# YAGI ANTENNA DESIGN FOR SIGNAL PHONE JAMMER

**[1]Y. FITRIYANI, [2]A.B. MUTIARA, [3]R. REFIANTI**

[1]Graduate Student, Department of Electrical Engineering, Gunadarma University, Indonesia

[2] Prof., Faculty of Computer Science and Information Technology, Gunadarma University, Indonesia

[3]Assist.Prof., Faculty of Computer Science and Information Technology, Gunadarma University, Indonesia

E-mail: [1]yi_jonke@yahoo.com , [2,3]{amutiara,rina@staff.gunadarma.ac.id}

**ABSTRACT**

Mobile phone is one of the most widely used today in mobile communications. This technology is very useful for communication but this raises several problems in a situation where silence is required such as in libraries, places of worship, classrooms and others. Mobile phone jammer is a device that used to block the incoming signal to a mobile phone from the base station. If the mobile phone jammer is turned on then it can't form the incoming or outgoing calls even sms. In this research, we designed a Yagi antenna (900MHz) to expand the range of jamming because Yagi has a great gain. Results of impedance by gamma match are 50.16 Omega. Obtained the value of VSWR Yagi is 1.46:1 and jamming distance that can be taken approximately 16 meters, It is different from the jamming distance of helical antenna on a mobile phone jammer itself is about 4 meters.

## 1. INTRODUCTION

Usage of mobile communications continues to increase along the development of variety of new technologies are increasingly powerful. One of the most widely used today is the mobile phone. Almost everyone has this advanced technology even some consumers have more than one mobile phone. Indeed, this technology is very useful for communication, but usage of mobile phone may also be disturbing.

Mobile phone jammer is a device used for blocking signals from the base station (BTS) that goes into the cellular phone. When tool is active, the jammer will automatically turn off cellular phones located nearby. The device is normally used in places where calls would be very disturbing because the silence is expected, such as in the classroom, worship places, meeting rooms, hospital, library and other places.

In this thesis we design an antenna for a GSM (Global System for Mobile) jammer which has frequency about the 890-960mhz. Jammer's antenna that commonly used helical antenna are included in directional antenna. This thesis create a Yagi antenna (directional) which could be adjusted the frequency to the range frequency of GSM. After that, we conduct the simulation by Supernec 2.9 and 4Nec2 to find out the values of parameters antenna. Doing experimental work using a mobile jammer with Yagi antenna that have been made and then discussed about the ratio of the coverage areas are achieved using a GSM jammer with helical antenna and yagi antenna that has been made.

## 2. JAMMER TECHNOLOGY

Previous, jammer technology used in military, jammer used to disrupt enemy communications and some jammers are also designed to foil the use of explosives to be detonated from a distance. But over time, some companies contracted to make the jammers and to sell these devices to private entities[6].

Mobile phone jammer blocking signal of mobile phone with transmits radio waves which has same frequency with mobile phone. It is causing interference communications between mobile phone and base station, so that mobile phone unusable.

Figure 1 shows jammer main block consists of power supply, IF section, RF Section and jamming signal. Function of IF section is to set the voltage to be received at the VCO in the RF section. RF section consists of power amplifier and the VCO (Voltage Controlled Oscillator), which serves to generate a signal.

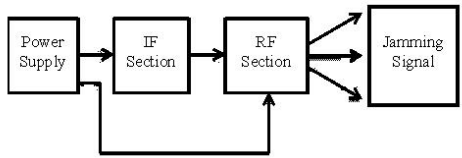

Figure1. Diagram Block of Jammer [1]

Five types of devices are known to have been developed (or being considered for development) for preventing mobile phones from ringing in certain specified locations [2]:

a. Type "A" Device. This type of devices transmits only a jamming signal and has very poor frequency selectivity, which leads to interference with a larger amount of communication spectrum than it was originally intended target.
b. Type "B" Device (Intelligent Cellular Disablers). Unlike type A, this type does not transmit an interfering signal on the control channels. The device basically works as a detector, and it capable to communicate with the cellular base station.
c. Type "C" Device (Intelligent Beacon Disablers). Like type B, this type does not transmit an interfering signal on the control channels. The device, when located in a specific silent room, function as a beacon and any compatible terminal is ordered to disable its ringer or disable its operation.
d. Type "D" Device (Direct Receive and Transmit Jammers). This type is similar to type A, but with a receiver, so that jammers is predominantly in receive mode and when the device detects the presence of a mobile phone in the silent room, it will intelligently choose to interact and block the cellular phone by transmitting jamming signal.
e. Type "E" Device (EMI Shield-Passive Jamming). This type is using technique EMI (Electromagnetic Interference) suppression to make a room into what I called a Faraday cage. Although labor intensive to construct, the Faraday cage essentially blocks, or greatly attenuates, virtually all electromagnetic radiation from entering or leaving the cage or in this case a target room.

A proper antenna is required to transmit a signal jamming. In order to have optimal power transfer, the antenna system must be matched to the transmission system.

On a mobile jammer antennas typically use a helix. This may be due to the helical antenna is more simple and has a smaller size so as to facilitate its use. Helical antenna is omnidirectional where the radiation pattern in all direction. The helix of mobile jammer has coverage jamming until 4 meters, so we will create a Yagi antenna for expand the coverage jamming because Yagi antenna has a great gain.

In Yagi antenna, increase of antenna alignment without power on all elements is expected. Elements that are not powered have parasitic character and receive signals from coupling driven element field. Basic construction is consists of driven added element parasitic, reflector (element parasitic), driven (element with powered) and director (also called array parasitic).

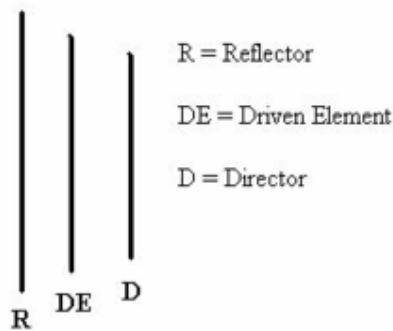

Figure 2. Elements of Yagi Antenna

## 3. SIMULATIONS OF YAGI ANTENNA

Firstly, to make an antenna, we must know the frequency as expected. In this research, we use GSM frequency are 900MHz and then calculate the wavelength value ($\lambda = 0.33$m).

In designing a Yagi antenna, length and spacing of each element has its own formulation. But there is no specific formula to make the best Yagi antenna on any band, however a lot of good Yagi design and can be tried was made[4]. Below is a Yagi antenna design according to some references as follows:

a. Constantine A. Balanis (Antenna Theory Analysis and Design)[7]
b. Yagi Antenna Design (NBS Technical Note 688)[10]
c. YC0PE by Ridwan Lesmana[13]

From three references above, we get parameters that suitable with formulation in respectively, see table 1.

Table 1. Design comparison Yagi antenna

| Elements | C.A.Balanis | | NBS TN 688 | | YC0PE | |
|---|---|---|---|---|---|---|
| | Length | Separation | Length | Separation | Length | Separation |
| R | 0.167 m | | 0.170 m | | 0.169 m | |
| | | 0.083 m | | 0.047 m | | 0.075 m |
| DE | 0.157 m | | 0.160 m | | 0.158 m | |
| | | 0.100 m | | 0.050 m | | 0.042 m |
| D1 | 0.147 m | | 0.140 m | | 0.150 m | |
| | | 0.110 m | | 0.043 m | | 0.058 m |
| D2 | 0.143 m | | 0.130 m | | 0.143 m | |
| | | 0.120 m | | 0.092 m | | 0.075 m |
| D3 | 0.140 m | | 0.120 m | | 0.137 m | |
| | | 0.130 m | | 0.130 m | | 0.092 m |
| D4 | 0.137 m | | 0.120 m | | 0.133 m | |

Software used to design a Yagi antenna and for the simulation is SuperNEC 2.9 and 4NEC2. After performing the simulation with SuperNEC and 4NEC2, for three different parameters, it can be seen the value of impedance, SWR and frequency along the greatest gain, as in the following table 2.

Table 2. Comparison simulation result

| Source | Impedance($\Omega$) | SWR | Max. Gain($dBi$) |
|---|---|---|---|
| C.A.Balanis | 83.4+j138 | 6.66 | -1.41 |
| NBS TN 688 | 24+j3.73 | 2.1 | 11.2 |
| YC0PE | 1.55+j25.6 | 40.8 | 6.53 |

From the table 2 above, it can be seen that the best parameters are according to NBS Technical Note 688 with a smaller value of SWR and gain the resulting greater. So these parameters are used for further design. To reduce the SWR value, we use a gamma match that is tuned and connected with the SWR analyzer.

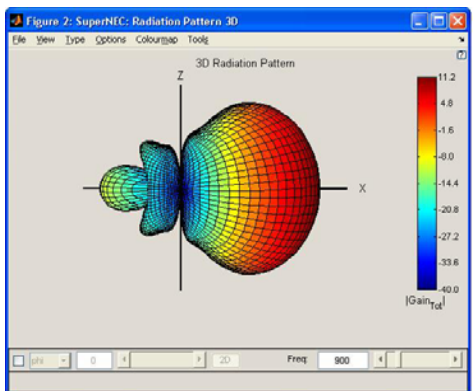

Figure 3. Radiation Pattern

Based on Figure 3, it is shown that radiation pattern has main lobe that is wide enough on the X axis, a back lobe and 2 sidelobe. From this radiation pattern, expected signal could jamming in long distance. Gain found at a frequency of 900 MHz is 11.2 dBi, where the gain is greatest at a frequency of 890-910 MHz and the smallest gain 10.9 dBi at a frequency of 960MHz. It means that Yagi antenna could be operation with the range frequency desired.

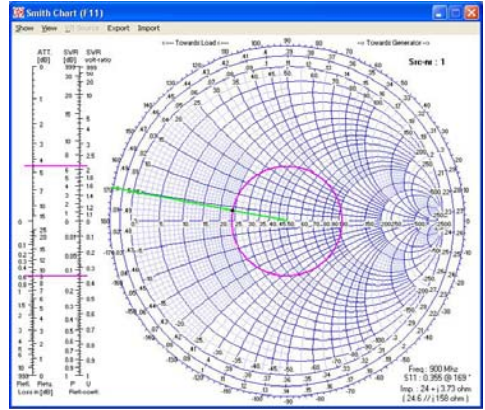

Figure 4. Smith Chart

Based on the smith chart of Figure4, can be seen that the impedance of the Yagi antenna that has been designed is 24 + j3.73$\Omega$. This certainly does not match the impedance used in this study is 50$\Omega$. To adjust the impedance, then the design will be used a gamma match as impedance technique that is placed on the driven element. Gamma match is one of impedance matching technique which is often used, because it is simpler.

## 4. DESIGN SPECIFICATION

The following are materials that are used to design a Yagi antenna, namely:

a. Aluminum Rod for elements, length 1 meter and a diameter of 5 mm.
b. Pipe boom 55 cm with a diameter of 2 cm
c. 5 Brackets with diameter 7/8 inch x 5 mm. This brackets is used to set elements (R, D1, D2, D3, and D4) on the pipe boom.
d. A bracket connector diameter 7/8 inch x 5 mm. This bracket is used to set DE and gamma match on the pipe boom.
e. N chassis connector is placed on the bracket connector.
f. Gammas match tube and Coax RG8 50$\Omega$.
g. N male and SMA male RG58 crimping 50$\Omega$.

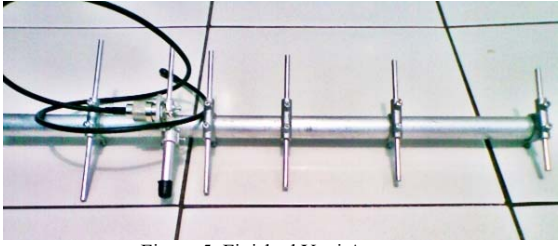

Figure 5. Finished Yagi Antenna

Cut the aluminum into the appropriate elements of the NBS consists of one reflektor, one driven and 4 director. See Figure 5.

Gamma match designed to match impedance. Open the outer layers of coaxial cable insert it into tube the gamma match, solder cable ends with a N chassis that has been placed on the connector bracket. We could see Figure 6 below.

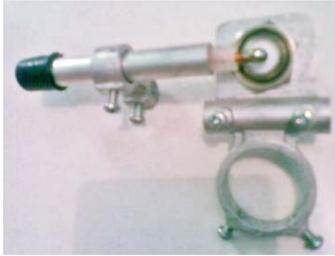

Figure 6. Gamma Match

## 5. RESULTS

Yagi antenna has been made should be measured its VSWR. The ideal of an antenna has VSWR = 1. In order for the ideal VSWR values, shifting the gamma rod up to the value of VSWR is better than the results of simulations conducted previously.

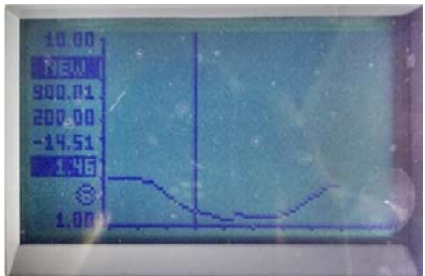

Figure 7. Display of SWR meters

Figure 7 shows that value of VSWR is 1.46 with return loss 14.51, while the antenna bandwidth is 130MHz.

From the sliding of the gamma match to get a small VSWR, impedance values can be calculated which is formed by the gamma match. Previously known to simulated impedance of Yagi antenna in free space without the gamma match are $24 + j3.73 \Omega$.

The diameter of the driven element and gamma match tubes 5 mm ($0.015\lambda=2a$) and 7.3 mm ($0.022\lambda=2a'$), respectively. The separation between the driven element and gamma rod is 17.2 mm ($0.052\lambda$). Then it can be calculated impedance values obtained as follows:

1. Determine the current division factor $\alpha$ by using equation (2.17), (2.18), and (2.19).

$$u = \frac{a}{a'} = \frac{0.011}{7.5 x 10^{-3}} = 1.5$$

$$v = \frac{s}{a'} = \frac{0.052}{7.5 x 10^{-3}} = 6.9$$

$$\alpha = \frac{\ln(v)}{\ln(v) - \ln(u)} = \frac{\ln(6.9)}{\ln(6.9) - \ln(1.5)} = 1.3$$

and the step-up ratio

$$(1+\alpha)^2 = (1+1.3)^2 = 5.29$$

2. The free-space impedance (without the gamma match) designate it as $Za$
$Za = 24 + j3.73$

3. Find the value of $Z_2$ by using equation (2.20).

$$Z_2 = (1+\alpha)^2 \frac{Z_a}{2} = (1+1.3)^2 x \left(\frac{24+j3.73}{2}\right) = 63.48 + 9.87\Omega$$

4. Determine impedance $Z0$ by using (2.21).

$$Z_0 = 276 \log_{10} \left(\frac{s}{\sqrt{aa'}}\right) = \left(\frac{0.052}{\sqrt{(0.011) x ((7.5 x 10^{-3}))}}\right) = 209$$

5. Normalize $z2$ by $Z0$, then:

$$z_2 = \frac{Z_2}{Z_0} = \frac{63.48 + j9.87}{209} = 0.3 + j0.05$$

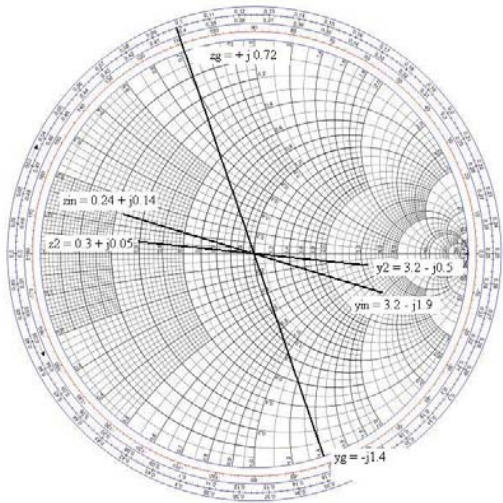

Figure 8. Smith Chart

6. On the Smith chart in figure8, locate $z_2$ and invert it to $y_2$.

$$y_2 = \frac{1}{z_2} = 3.2 - j0.5$$

7. On the Smith chart, locate $zg = 0+j0$ and advance it toward the generator a distance $0.099\lambda$ to obtain, $zg = 0 + j0.72$
8. From the Smith chart,

$$y_g = \frac{1}{z_g} = -j1.4$$

9. Add $y_2$ and $y_g$

$$y_{in} = y_2 + y_g = 3.2 - j1.9$$

which is located in the Smith chart.

10. Inverting $yin$ on the Smith chart to $zin$ gives

$$z_{in} = 0.24 + j0.14$$

11. Unnormalizing $zin$ by $Z0 = 209$

$$Z_{in} = z_{in} x Z_0 = (0.24 + j0.14) \times 209 = 50.16 + j29.26\Omega$$

12. The capacitance should be

$$C = \frac{1}{2\pi f_0 (29.26)} = \frac{1}{2\pi (900 x 10^6)(29.26)} = 6.04 x 10^{-12} = 6pF$$

## 6. ANALYSIS

From Figure 8 we got the radiation of jamming using helical antenna below. Maximum distance of jamming is 4.6 at 190 degree. From measurement that has been done, it can be seen that the average of area jamming is 3 to 4 meters.

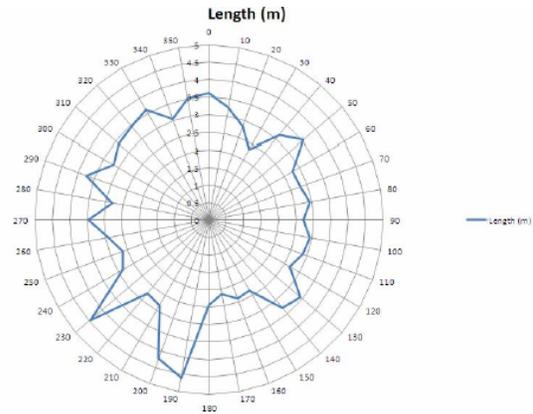

Figure 9. Radiation Pattern of Helical Antenna

The maximum distance of 16.72 meters jamming largest found at an angle 10 degrees and a minimum distance of 3.12 meters at an angle of 120 degrees. Distance at every angle is di_erent, this is because the characteristics of Yagi antenna as directional antenna which is the antenna with the radiation in one direction. Direction of the antenna radiation focused in one direction so that the resulting gain Yagi is greater.

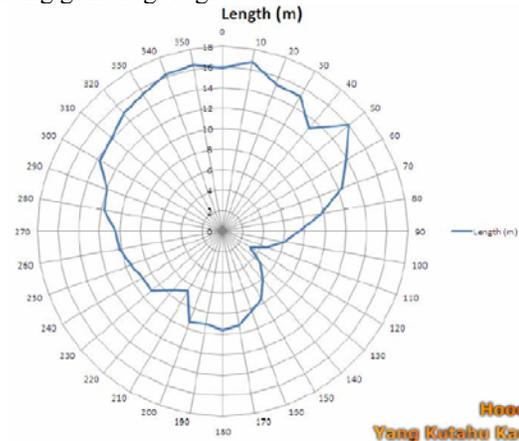

Figure 10. Radiation Pattern of Yagi Antenna

## 7. CONCLUSION AND FUTURE WORK

1. VSWR of made Yagi antenna is 1.46:1 with RL = 14.51 dB.
2. Bandwidth that achieved by Yagi antenna is about 130 MHz, this is more than range of GSM.
3. Impedance matching using gamma match achieved ± 50.16Ω and the capacitance is ± 6 pF.
4. Designed Yagi antenna can jamming a mobile phone up to 16 meters with a fourfold increase compared with helical antenna.

For future work, this research can be developed by creating a variety of other antennas for mobile phone jammer with higher frequencies such as 3G (1800-1900MHz) and DCS (2100-2200MHz).